\documentstyle[preprint,aps,graphicx]{revtex}

\def\err#1#2{$\stackrel{\scriptstyle +#1}{\scriptstyle -#2}$}
\begin{document}
\title{Constraints on Electron-quark Contact Interactions}
\author{Kingman Cheung
\footnote{Plenary talks presented at the IVth International Workshop on 
Particle Physics Phenomenology, National Sun yat-sen University, Taiwan R.O.C.,
June 18--21 1998
and at the D0 Collaboration meeting, Davis, CA, March 1998; and
in a parallel session at PASCOS-98, Northeastern University, Boson MA,
March 1998.}
}
\address{Department of Physics, University of California, Davis CA 95616}
\maketitle
\begin{abstract}
In this talk, I summarize a global analysis on electron-quark contact 
interactions 
using the updated NC DIS data at HERA, Drell-yan production at
the Tevatron, total hadronic cross sections at LEP, atomic parity
violation measurement, low energy $e$-nucleon scattering data, and the
$\nu$-nucleon scattering data.
The global data do not show any evidence for contact interactions.  Thus,
we obtain limits of 8--15 TeV on the compositeness scale, which are 
significantly better than those published by each individual experiment.
\end{abstract}

\section{Introduction}

Four-fermion contact interaction is not something new, but was proposed 
decades ago by Fermi to account for the nuclear beta decay. 
The interaction is represented by
\begin{equation}
{\cal L} \sim G_F \left( \bar e \gamma^\mu (1-\gamma^5) \nu \right)\;
                  \left( \bar u \gamma_\mu (1-\gamma^5) d \right) \;
\end{equation}
where $G_F$ is the fermi constant with dimension (mass)$^{-2}$.
This interaction is not renormalizable because amplitudes grow
indefinitely with the energy scale if $G_F$ is kept constant.  
It was only until
60's that the electroweak theory was proposed.  The four-fermion contact 
interaction was then replaced by exchange of weak gauge bosons and
$G_F$ replaced by the $W$ boson propagator: $G_F \to 1/(p^2 - m_W^2)$.
The weak gauge bosons were only
discovered later when the energy scale reached the hundred GeV level.
In the above history we learn a couple of lessons:
(i) the existence of four-fermion contact interactions is a signal of
new physics beyond the existing standard theory;
(ii) the exact nature of new physics is unknown at the low energy 
scale.  Only when the energy scale is high enough can the nature of new 
physics be probed.

The purpose of this analysis \cite{ours} is to examine the data from
current 
accelerator experiments to see if there is any sign of contact interactions.
If so it is a signal of new physics; if not we put limits
on the compositeness scale $\Lambda$.
Four-fermion contact interactions have been searched in recent high energy
experiments: (i) the $qqqq$ contact interaction studied at the Tevatron
by CDF \cite{cdf}; (ii) $eeqq$ interactions at LEP \cite{lep}, 
HERA \cite{hera-cont}, Drell-yan production at the Tevatron \cite{cdf-dy,D0}, 
and low energy $e$-Nucleon scattering experiments \cite{eN}.
We concentrate on $eeqq$ contact interactions.
In particular, the neutral current (NC) deep-inelastic scattering (DIS)
data collected by H1 \cite{H1} and ZEUS \cite{zeus} at HERA  between 1994--96
showed a significant excess in cross section in the high-$Q^2$ region,
which aroused an enormous amount of phenomenological activities.  One of
the explanations is the $eeqq$ contact interaction at the scale of 3 TeV.
However, the data collected in 1997 alone agreed very well with the SM and, 
therefore, the logical explanation for the excess in 1996 was statistical
fluctuation.  The overall result of the combined 94-97 data is as follows
\cite{hera-talks}:
(i) data by ZEUS agreed with the SM expectation up to $Q^2 \sim 30000 \,
{\rm GeV}^2$ while there are 2 events at $Q^2>35000\,{\rm GeV}^2$, where
only $0.29\pm0.02$ is expected;
(ii) data by H1 only showed a slight deviation above the SM for 
$Q^2>15000\,{\rm GeV}^2$ and the excess events in the mass window of 200 GeV
are now much less significant.
Although the outcome is somewhat discouraging for searching for new physics, 
we can, however, use the data to constrain new physics.
The objective here is to constrain the $eeqq$ contact interactions
using the global data, which include:
(i) NC DIS data at HERA \cite{hera-talks}, 
(ii) Drell-yan production at the Tevatron \cite{cdf-dy},
(iii) total hadronic cross sections at LEP and the left-right asymmetry 
at SLD \cite{lep},
(iv) atomic physics parity violation measurement \cite{apv}
on $^{133}Cs$,
(v) low energy $e$-N scattering experiments \cite{eN}, and 
(vi) low energy $\nu$-N scattering experiment by CCFR \cite{ccfr}.
We shall obtain fits of parameters of $eeqq$ contact interactions
and finally the limits on the compositeness scale $\Lambda$.  

In this write-up, we shall summarize the analysis in Ref. \cite{ours}.
The results presented here, however, use the more updated data since 
Ref. \cite{ours}.

\section{Parameterization}

The conventional effective Lagrangian of $e e q q$ contact
interactions has the form
\begin{eqnarray}
L_{NC} &=& \sum_q \Bigl[ \eta_{LL}
\left(\overline{e_L} \gamma_\mu e_L\right)
\left(\overline{q_L} \gamma^\mu q_L \right)
+ \eta_{RR} \left(\overline{e_R}\gamma_\mu e_R\right)
                 \left(\overline{q_R}\gamma^\mu q_R\right) \nonumber\\
&& \quad + \eta_{LR} \left(\overline{e_L} \gamma_\mu e_L\right)
                             \left(\overline{q_R}\gamma^\mu q_R\right)
+ \eta_{RL} \left(\overline{e_R} \gamma_\mu e_R\right)
\left(\overline{q_L} \gamma^\mu q_L \right) \Bigr] \,, \label{effL}
\end{eqnarray}
where eight independent coefficients $\eta_{\alpha\beta}^{eu}$ and
$\eta_{\alpha\beta}^{ed}$ have dimension (TeV)$^{-2}$ and are conventionally
expressed as $\eta_{\alpha\beta}^{eq} = \epsilon g^2 /\Lambda_{eq}^2$,
with a fixed $g^2=4\pi$.
The sign factor $\epsilon= \pm 1$ allows for either constructive or destructive
interference with the SM $\gamma$ and $Z$ exchange amplitudes and
$\Lambda_{eq}$ represents the mass scale of the exchanged new
particles, with coupling strength $g^2/4\pi=1$. A coupling of this
order is expected in substructure models and therefore $\Lambda_{eq}$ is
sometimes called the ``compositeness scale''.

%\subsection{$SU(2)\times U(1)$ symmetry and universality }

Left-handed electrons and quarks belong to SU(2) doublets $L=(\nu_L,e_L)$
and $Q=(u_L,d_L)$ and thus from SU(2) symmetry
one expects relations between contact terms involving left-handed
$u$ or $d$ quarks; similarly, contact terms for left-handed electrons and
neutrinos should be related. 
We start with the most general $SU(2)\times U(1)$ invariant contact term
Lagrangian,
\begin{eqnarray}
{\cal L}_{SU(2)}&=&
\eta_1 \Bigl(\overline L\gamma^\mu L\Bigr) \Bigl(\overline Q\gamma_\mu Q\Bigr)
+
\eta_2 \Bigl(\overline L\gamma^\mu T^aL\Bigr) \Bigl(\overline Q\gamma_\mu T^a
Q\Bigr) \nonumber \\
&& + 
\eta_3 \Bigl(\overline L\gamma^\mu L\Bigr)
\Bigl(\overline{u_R}\gamma_\mu
u_R\Bigr)
+ \eta_4 \Bigl(\overline L\gamma^\mu L\Bigr)
\Bigl(\overline{d_R}\gamma_\mu
d_R\Bigr) \nonumber \\
&&+ 
\eta_5 \Bigl(\overline{e_R}\gamma^\mu e_R\Bigr) \Bigl(\overline Q\gamma_\mu
Q\Bigr) +
\eta_6 \Bigl(\overline{e_R}\gamma^\mu e_R\Bigr) \Bigl(\overline{u_R}\gamma_\mu
u_R\Bigr) \nonumber\\
&& + \eta_7 \Bigl(\overline{e_R}\gamma^\mu e_R\Bigr)
\Bigl(\overline{d_R}\gamma_\mu
d_R\Bigr)\; .
\label{effLsu2}
\end{eqnarray}
By expanding the $\eta_5$ term we have 
\begin{equation}
\eta^{eu}_{RL}=\eta_5=\eta^{ed}_{RL} \; .
\label{su2releR}
\end{equation}
In addition, the four neutrino and the lepton couplings are related by SU(2),
\begin{equation}
\eta^{\nu u}_{LL}  = \eta^{ed}_{LL}\; , 
\eta^{\nu d}_{LL}  = \eta^{eu}_{LL}\; , 
\eta^{\nu u}_{LR}  = \eta^{eu}_{LR}\; , 
\eta^{\nu d}_{LR}  = \eta^{ed}_{LR}\; . 
\label{su2relnu}
\end{equation}
In our analysis, the relations of Eqs.~(\ref{su2releR}) and (\ref{su2relnu})
are only used when neutrino scattering data are included in the analysis. Even
though we expect that $SU(2)\times U(1)$ will be a
symmetry of the renormalizable interactions which ultimately manifest
themselves
as the contact terms of Eq.~(\ref{effL}), electroweak symmetry breaking may
break the degeneracy of SU(2) multiplets of new, heavy quanta whose exchanges
give rise to (\ref{effL}). This would result in a violation of the relations
of Eqs.~(\ref{su2releR}) and (\ref{su2relnu}). One example is the exchange
of the stop $\tilde t_1$, $\tilde t_2$, and the sbottom $\tilde b_L, \tilde
b_R$
in $R$-parity violating SUSY models. The large top-quark mass may lead to
substantial splitting of the masses of these squarks which could easily
lead to violations by up to a factor of two of SU(2) relations such as
$\eta^{\nu d}_{LR}=\eta^{ed}_{LR}$.

Because of severe
experimental constraints on intergenerational transitions like
$K\to\mu e$
we restrict our discussion to first generation contact terms. Only
where required by particular data (e.g. the muon sample of Drell-yan
production at the Tevatron) will we assume universality of contact terms 
between $e$ and $\mu$.

\section{Global Data}

The global data used in this analysis have been described in Ref. \cite{ours}.
Here we only list those that have been updated since then.

\subsection{HERA data}

The 1997 data alone by H1 and ZEUS agreed very well with the SM expectation,
though the combined 1994--1997 data still showed an excess of cross section
at high $Q^2$.  The significance of excess is far less severe now.  
We use the $Q^2$ distribution presented in the 1998 spring conferences
\cite{hera-talks}.  Note that
using $Q^2$ distribution will not reduce appreciably the sensitivity to 
contact interactions than using the $x$-$y$ distribution, because $x$ 
distribution is not sensitive to contact interactions (unlike the narrow-width
leptoquark model) but $y$ is somewhat sensitive to it.  
The updated data are tabulated in Table \ref{table1}.

\begin{table}[ht]
\caption[]{
\label{table1}
\small The measured number of events as a function of 
$Q^2_{\rm min}$ at HERA.}
\medskip
\centering
\begin{tabular}{|ccl||ccl|}
\hline
\multicolumn{3}{|c||}{ZEUS (${\cal L}=46.60\, {\rm pb}^{-1}$) }& 
\multicolumn{3}{c|}{H1 (${\cal L}=37.04\, {\rm pb}^{-1}$) } \\
\hline
\underline{$Q^2_{\rm min} \,({\rm GeV}^2)$} & \underline{$N_{\rm obs}$}
 & \underline{$N_{exp}$} & 
\underline{$Q^2_{\rm min} \,({\rm GeV}^2)$} & \underline{$N_{\rm obs}$} &
\underline{$N_{exp}$} \\
2500 & 1817 & 1792$\pm93$ &  2500 & 1297 & 1276$\pm98$ \\ 
5000 & 440  & 396$\pm24$ &  5000  & 322  & 336$\pm29.6$ \\
10000 & 66  & 60$\pm4$ &  10000  & 51   & 55.0$\pm6.42$ \\
15000 & 20  &  17$\pm2$ & 15000  & 22   & 14.8$\pm2.13$ \\
35000 & 2   & 0.29$\pm0.02$ & 20000 & 10   & 4.39$\pm0.73$ \\
      &     &      & 25000 & 6    & 1.58$\pm0.29$ \\
\hline
\end{tabular}
\end{table}

\subsection{Drell-yan Production}

Since our previous analysis, in which we used the plotted data on a CDF graph,
CDF \cite{cdf-dy} has published the observed number of events in bins of 
invariant mass of the lepton pair.  The data are given in Table \ref{table2}.
In this CDF paper, they obtained limits on the compositeness scale using
only the CDF data, in the order of a few TeV
\footnote{
D0 Collaboration \cite{D0} has also recently published a paper on
``Search for Quark-lepton Compositeness using the Drell-Yan process at D0''.
}.
At the end, we shall obtain limits significantly better than these limits.

\begin{table}[ht]
\caption[]{ \label{table2}
\small The electron and muon samples of Drell-yan production
in each mass bin by CDF.}
\medskip
\centering
\begin{tabular}{|c|cc||cc|}
\hline
& 
\multicolumn{2}{c||}{$e^+ e^-$} & \multicolumn{2}{c|}{$\mu^+ \mu^-$} \\
\hline
\underline{$M_{\ell\ell}$} & \underline{$N_{\rm obs}$}
 & \underline{$N_{\rm exp}$} & \underline{$N_{\rm obs}$}
 & \underline{$N_{\rm exp}$}  \\
50--150  &  2581 & 2581 & 2533 & 2533 \\ 
150--200 &  8    & 10.8 &   9  & 9.7  \\
200--250 &  5    & 3.5  &   4  & 3.2  \\
250--300 &  2    & 1.4  &   2  & 1.3  \\
300--400 &  1    & 0.97 &   1  &  0.94 \\
400--500 &  1    & 0.25 &   0  &  0.27 \\
500--600 &  0    & 0.069 &  0  & 0.087 \\
\hline
\end{tabular}
\end{table}

\subsection{LEP}

The LEP collaborations have published new measurements of total hadronic
cross sections at $\sqrt{s}=130, 172$, and 183 GeV.  The data that we used
are shown in Table \ref{table3}.

\begin{table}[ht]
\caption[]{
\label{table3}
\small Total hadronic cross sections $\sigma_{\rm had}$ measured
by the LEP collaborations.}
\medskip
\centering
\begin{tabular}{|ccc|}
\hline
$\sqrt{s}$ (GeV) &  $\sigma_{\rm had}$  & $\sigma_{\rm SM}$ \\
\hline
\multicolumn{3}{|c|}{ALEPH} \\
\hline
130 & 79.5 $\pm$ 4.14 &  77.16 \\
136 & 64.5 $\pm$ 3.85 &  62.52 \\
183 & 23.6 $\pm$ 0.73 & 23.05 \\
\hline
\multicolumn{3}{|c|}{DELPHI}\\
\hline
130.2 & 82.2 $\pm$ 5.2 &  83.1 \\
136.2 & 65.9 $\pm$ 4.7 &  67.0 \\
161.3 & 40.2 $\pm$ 2.1 &  34.8 \\ 
172.1 & 30.6 $\pm$ 2.0 &  28.9 \\
\hline
\multicolumn{3}{|c|}{L3}\\
\hline
130.3 & 81.8 $\pm$ 6.4 &  78 \\
136.3 & 70.5 $\pm$ 6.2 &  63 \\
140.2 & 67   $\pm$ 47  &  56 \\
161.3 & 37.3 $\pm$ 2.2 &  34.9 \\ 
170.3 & 39.5 $\pm$ 7.5 &  29.8 \\
172.3 & 28.2 $\pm$ 2.2 &  28.9 \\
\hline
\multicolumn{3}{|c|}{OPAL} \\
\hline
130.25 & 64.3 $\pm$ 5.1 &  77.6 \\
136.22 & 63.8 $\pm$ 5.2 &  62.9 \\
161.34 & 35.5 $\pm$ 2.2 &  33.7 \\ 
172.12 & 27.0 $\pm$ 1.9 &  27.6 \\
\hline
\end{tabular}
\end{table}

\section{Fits and Limits}

\begin{table}[th]
\caption[]{\label{table4}
\small
The best estimate of the $\eta_{\alpha\beta}^{eq}$ parameters when various
data sets are added successively.  In the last column when the $\nu$-N
data are included the $\eta_{L\beta}^{\nu q}$ are given in terms of
$\eta_{L\beta}^{eq}$ by Eq.~(\protect\ref{su2relnu}) and
we assume $\eta_{RL}^{eu}=\eta_{RL}^{ed}$ in the last column.
}
\medskip
\centering
\begin{tabular}{|l|c|c|c|c|c|}
\hline
 & HERA only & HERA+APV & HERA+APV & HERA+APV  &  HERA+DY+APV \\
 &           & +eN      & +eN+DY   & +eN+DY+LEP &+eN+LEP+$\nu$N \\
\hline
$\eta_{LL}^{eu}$ & 2.04\err{3.97}{5.26}  
  & 2.25\err{2.29}{3.63}  & 0.22\err{0.67}{0.57}
  & 0.049\err{0.63}{0.43}  & -0.046\err{0.62}{0.38} \\
$\eta_{LR}^{eu}$ & -4.30\err{4.30}{0.78} 
  & -2.77\err{3.20}{1.70}  & 0.60\err{0.51}{0.66}
 & 0.76\err{0.38}{0.63}  & 0.76\err{0.35}{0.68} \\
$\eta_{RL}^{eu}$ & -1.75\err{3.75}{2.59} 
  & -3.53\err{2.91}{0.90}  & -0.004\err{0.72}{0.72}
  & 0.042\err{0.73}{0.75}  & 0.13\err{0.73}{0.77} \\
$\eta_{RR}^{eu}$ & 2.62\err{4.28}{5.36}  
 & 2.23\err{1.77}{3.41} & 0.040\err{0.66}{0.62}
 & -0.051\err{0.68}{0.57} & -0.091\err{0.71}{0.56} \\
$\eta_{LL}^{ed}$ & -1.71\err{7.87}{6.77} 
 & -2.22\err{5.62}{4.55}  & 0.25\err{1.64}{1.72}
 & 0.39\err{0.76}{0.92}  & 0.11\err{0.82}{0.53} \\
$\eta_{LR}^{ed}$ & -0.011\err{4.85}{4.48} 
 & -0.95\err{3.76}{3.47}  & 1.65\err{1.39}{2.79}
 & 0.79\err{1.33}{2.08}  & 0.36\err{1.16}{1.82} \\
$\eta_{RL}^{ed}$ & -1.86\err{4.86}{4.38} 
 & -0.95\err{3.87}{3.23}  & 1.97\err{1.30}{2.74}
 & 1.11\err{1.27}{2.02}  & $=\eta_{RL}^{eu}$ \\
$\eta_{RR}^{ed}$ & -2.28\err{7.87}{7.22} 
 & -1.61\err{5.59}{4.85}  & 0.55\err{1.60}{1.73}
 & 0.73\err{0.85}{1.03}  & 0.86\err{0.60}{1.16} \\
\hline
\hline
HERA    & 7.57 & 7.86  & 12.10 & 12.34  & 12.73  \\
APV	&      & 0.00  & 0.00  &  0.001 & 0.001  \\
eN      &      & 0.46  & 0.47  &  0.46  & 0.51  \\
DY      &      &       & 4.40  &  4.38  & 4.39 \\
LEP     &      &       &       &  23.57  & 23.49 \\
$\nu$N  &      &       &       &        &  0.00   \\
\hline
\hline
Total $\chi^2$ & 7.57 & 8.32   & 16.97 &  40.75 & 41.12  \\
\hline
SM $\chi^2$ & 17.27 &  20.27 & 24.55 & 51.20 & 51.21 \\
\hline
SM d.o.f.   & 11    &  16     & 28     & 47     & 48 \\
\hline
\end{tabular}
\end{table}

\begin{figure}[ht]
\leavevmode
\begin{center}
\includegraphics[width=3.2in]{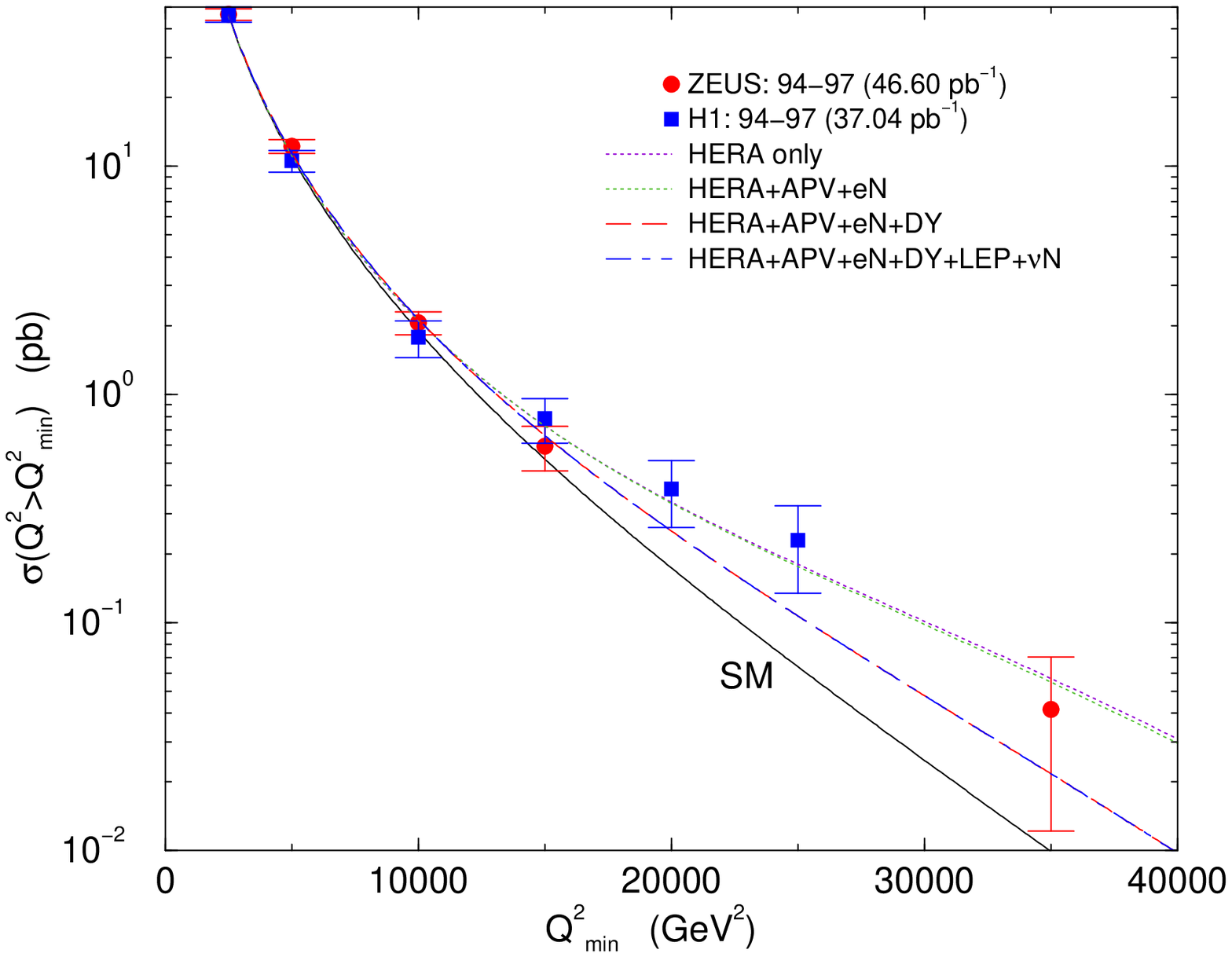}
\includegraphics[width=3.05in]{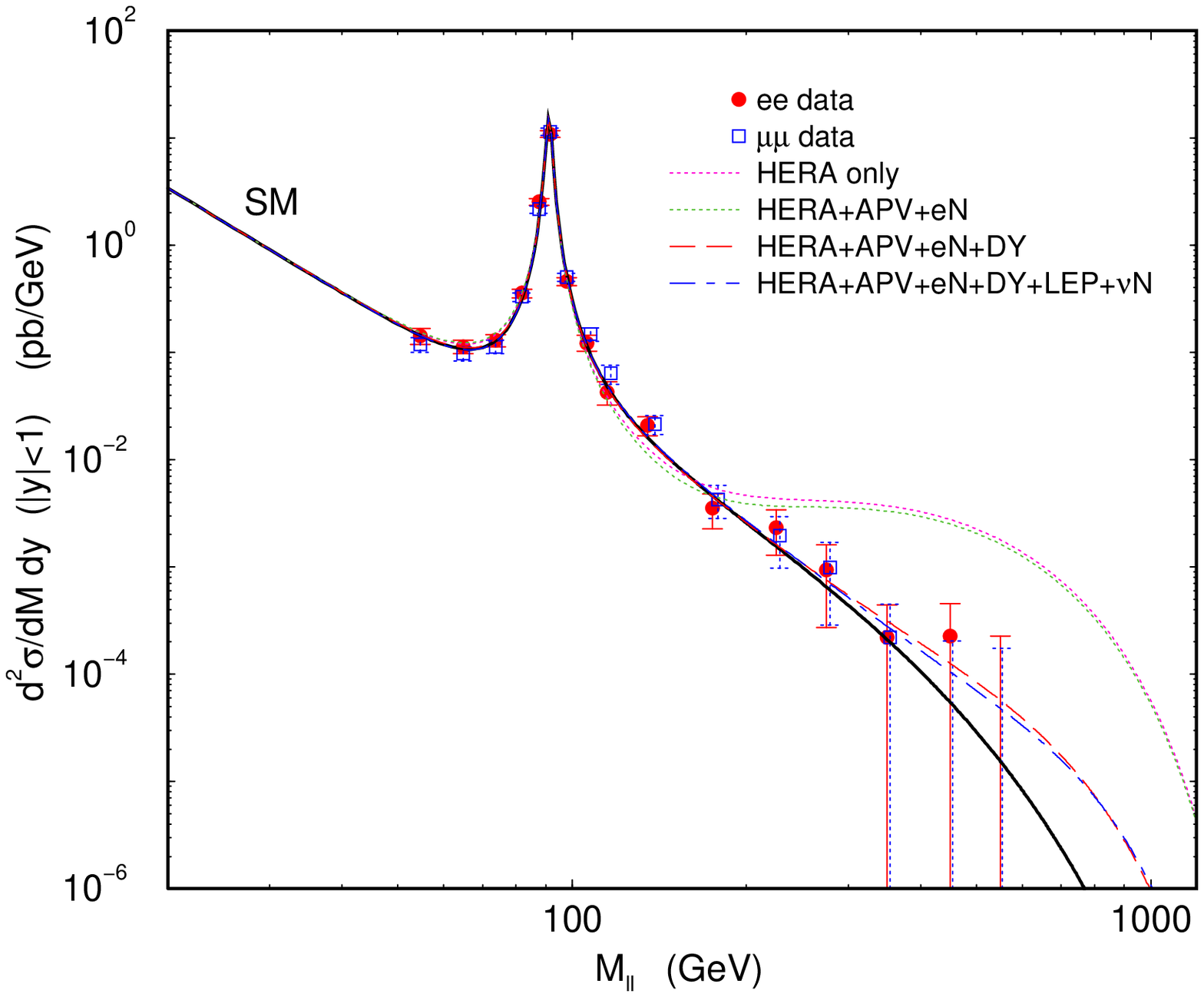}
\end{center}
\caption{\label{fig1} \small 
(a) The cumulated cross section $\sigma(Q^2 > Q^2_{\rm min})$ 
at HERA as a function of $Q^2_{\rm min}$.  The 94--97 H1 and ZEUS data, and
curves of fits to various data sets are shown,   
(b) the differential cross section $d^2\sigma/dM dy$ for Drell-yan 
production at the Tevatron.  }
\end{figure}

The fits of contact parameters are obtained by minimizing the $\chi^2$ of
the data sets.  In order to see how each data set affects the fit, we
obtain the fit with each data set added one at a time.
The fits with various combinations of data sets are shown in 
Table \ref{table4}.
Two important observations are offered as follows.
(i) When the Drell-yan data are added to ``HERA$+$APV$+e$N'', the fitted
parameters change dramatically and so are the $\chi^2$'s of each data set.
This can be understood as follows.  The HERA data actually favor non-zero
contact parameters (especially the last entries of ZEUS and H1 data): see
Fig. \ref{fig1}(a).  However, this fit of contact parameters very much
contradicts the Drell-yan data: see Fig. \ref{fig1}(b).  Therefore, 
when DY data are taken into account, the fit changes drastically.  The curves
of the fit with all data sets are also shown in Fig. \ref{fig1}.
(ii) The goodness of the fits is indicated by the $\chi^2$ per degree of 
freedom (d.o.f.). 
The $\chi^2$/d.o.f. ($\chi^2_{\rm cont.}$/d.o.f.=1.003) 
of contact interactions is very close to that
of the SM ($\chi^2_{\rm SM}$/d.o.f.=1.067) for the last column in 
Table \ref{table4}.
For the second last column in Table \ref{table4}, 
$\chi^2_{\rm cont.}$/d.o.f.= 1.045
and $\chi^2_{\rm SM}$/d.o.f.= 1.089.

In view of these, we conclude that the global data do not show any sign 
of contact interactions.  Thus, we can derive 95\% CL limits on the 
compositeness scale, below which the contact interaction is ruled out.
The limits on $\Lambda_{\pm}$ are listed in Table \ref{table5}--\ref{table7}.
In Table \ref{table5}, for each chirality coupling 
considered the others are put to
zero. The limits on $\Lambda$ obtained range from 8--15 TeV, which improve
significantly from each individual experiment \cite{lep,hera-cont,cdf-dy,D0}.
We also calculate the limits on the compositeness scale when some symmetries
on contact terms are considered, as shown in Tables \ref{table6} and 
\ref{table7}.
$VV$ stands for vector-vector:
$\eta_{LL}=\eta_{LR}=\eta_{RL}=\eta_{RR}=\eta_{VV}$, while
$AA$ stands for axial-vector-axial-vector:
$\eta_{LL}=-\eta_{LR}=-\eta_{RL}=\eta_{RR}=\eta_{AA}$.
These limits, in general, are 
not as strong as those in the previous table because the additional symmetry
automatically satisfies the parity violation experiments: APV and $e$-N.

\begin{table}[th]
\caption[]{\label{table5}
\small
The best estimate on $\eta_{\alpha\beta}^{eq}$ and the 95\% CL limits on the
compositeness scale $\Lambda_{\alpha\beta}^{eq}$,
where $\eta_{\alpha\beta}^{eq}=4\pi\epsilon/(
\Lambda_{\alpha\beta\epsilon}^{eq})^2$.
When one of the $\eta$'s is considered the others are set to zero.
SU(2) relations are assumed and $\nu$N data are included.
}
\medskip
\centering
\begin{tabular}{|cc|cc|}
\hline
& & \multicolumn{2}{c|}{95\% CL Limits} \\
Chirality  ($q$) &  $\eta\;\;$ (TeV$^{-2}$)  &  
$\Lambda_+$ (TeV) & $\Lambda_-$ (TeV) \\
\hline
\hline
LL($u$)  & $0.026 \pm 0.056$ &  9.9  & 11.6 \\
LR($u$)  & $0.11  \pm 0.079$ &  7.3  & 11.1 \\
RL($u$)  & $-0.043\pm 0.038$ &  15.5 & 10.8 \\
RR($u$)  & $-0.12 \pm 0.078$ &  11.5 & 7.0 \\
LL($d$)  & $0.072 \pm 0.060$ &  8.5  & 12.5 \\
LR($d$)  & $0.079 \pm 0.072$ &  7.8  & 11.2\\
RR($d$)  & $-0.064\pm 0.072$ &  10.9 & 8.1 \\
\hline
\end{tabular}
\end{table}

\begin{table}[th]
\caption[]{\label{table6}
\small
The best estimate on $\eta^{eq}$ for the minimal setting, $VV,AA$, and
SU(12), and the corresponding 95\% CL limits on the
compositeness scale $\Lambda$, where $\eta=4\pi\epsilon/
(\Lambda_{\epsilon})^2$.
When one of the $\eta$'s is considered the others are set to zero.
Here we do not use SU(2) relations nor do we include the $\nu$N data.
}
\medskip
\centering
\begin{tabular}{|cc|cc|}
\hline
& & \multicolumn{2}{c|}{95\% CL Limits} \\
Chirality  ($q$) &  $\eta\;\;$ (TeV$^{-2}$) &
$\Lambda_+$ (TeV)  & $\Lambda_-$ (TeV) \\
\hline
\hline
$\eta_{LR}^{eu}=\eta_{RL}^{eu}$  & 0.41 \err{0.23}{0.27}  & 4.0 & 6.1 \\
$\eta_{LR}^{ed}=\eta_{RL}^{ed}$  & $-1.19$\err{0.37}{0.30}&2.3 & 2.8  \\
\hline
$\eta_{VV}^{eu}$   & $-0.064$\err{0.090}{0.089} &  9.3 & 7.6 \\
$\eta_{VV}^{ed}$   & 0.38\err{0.18}{0.21} &  4.4 & 4.9 \\
\hline
$\eta_{AA}^{eu}$  & $-0.30$\err{0.12}{0.11} & 9.4  &  5.1 \\
$\eta_{AA}^{ed}$  & 0.31 \err{0.15}{0.16}   & 4.8  &  7.5 \\
\hline
$\eta_{LL}^{eu}=-\eta_{LR}^{eu}$  & $-0.47$\err{0.19}{0.18} &6.5& 4.1 \\
$\eta_{RL}^{eu}=-\eta_{RR}^{eu}$  & 0.54\err{0.19}{0.21}  & 3.9 & 5.8 \\
$\eta_{LL}^{ed}=-\eta_{LR}^{ed}$  & 0.54\err{0.24}{0.27}  & 3.7 & 4.8 \\
$\eta_{RL}^{ed}=-\eta_{RR}^{ed}$  & $-0.62$\err{0.30}{0.26} & 3.6 & 3.5 \\
\hline							    
\end{tabular}						    
\end{table}

\begin{table}[th]
\caption[]{\label{table7}
\small
Same as the last Table but with a further condition: $\eta^{eu}=\eta^{ed}$.
Here $q=u=d$.
}
\medskip
\centering
\begin{tabular}{|cc|cc|}
\hline
& & \multicolumn{2}{c|}{95\% CL Limits} \\
Chirality  ($q$) &  $\eta\;\;$ (TeV$^{-2}$) &
 $\Lambda_+$ (TeV)  & $\Lambda_-$ (TeV) \\
\hline
\hline
$\eta_{LR}^{eq}=\eta_{RL}^{eq}$  & 0.46 \err{0.25}{0.33} & 3.9  & 5.5 \\
\hline
$\eta_{VV}^{eq}$   & $-0.026$\err{0.15}{0.13}  &  5.3  & 6.9 \\
\hline
$\eta_{AA}^{eq}$  & $-0.40$\err{0.17}{0.15}  & 4.3   & 4.5 \\
\hline
$\eta_{LL}^{eq}=-\eta_{LR}^{eq}$  & $-0.61$\err{0.25}{0.21} &3.4 & 3.7 \\
$\eta_{RL}^{eq}=-\eta_{RR}^{eq}$  & 0.65\err{0.21}{0.25} & 3.6 & 3.3\\
\hline
\end{tabular}
\end{table}

In conclusion, the global data have been examined and do not 
support the existence of $eeqq$ contact interactions with the compositeness
scale upto 8--15 TeV.  Although the 1994--97 the NC DIS data at HERA favor
a slightly non-zero contact interaction, the other data, especially the
Drell-yan data at the Tevatron and the atomic parity violation measurement,
severely constrain it.  Finally, the limits on the compositeness scale obtained
in this analysis are significantly better than the results published by
each individual experiment.  We urge others to use the limits of $\Lambda$
obtained in this analysis.
The above analysis can also be applied in a straight-forward fashion to other 
new physics such as $Z'$ and leptoquark models.

I would like to thank Vernon Barger, Karou Hagiwara, and Dieter Zeppenfeld for
collaboration.

\end{document}